\documentclass[sigconf]{acmart}

\usepackage{enumitem}
\usepackage{color}
\usepackage{subcaption}
\newcommand{\delete}[1]{}
\newcommand{\add}[1]{\textcolor{black}{#1}}
\newcommand{\addfinal}[1]{\textcolor{black}{#1}}
\AtBeginDocument{%
  }

\copyrightyear{2025}
\acmYear{2025}
\setcopyright{cc}
\setcctype{by}
\acmConference[CHI '25]{CHI Conference on Human Factors in Computing Systems}{April 26-May 1, 2025}{Yokohama, Japan}
\acmBooktitle{CHI Conference on Human Factors in Computing Systems (CHI '25), April 26-May 1, 2025, Yokohama, Japan}\acmDOI{10.1145/3706598.3714162}
\acmISBN{979-8-4007-1394-1/25/04}

\begin{document}



\author{Jian Zhang}
\affiliation{%
  \institution{University of Melbourne}
  \city{Melbourne}
  \state{VIC}
  \country{Australia}}
  \orcid{0009-0009-1619-8297}\email{jianzhang10@student.unimelb.edu.au}

\author{Wafa Johal}
\affiliation{%
  \institution{University of Melbourne}
  \city{Melbourne}
  \state{VIC}
  \country{Australia}
}
\orcid{0000-0001-9118-0454}
\email{wafa.johal@unimelb.edu.au}

\author{Jarrod Knibbe}
\affiliation{%
  \institution{The University of Queensland}
  \city{St Lucia}
  \state{QLD}
  \country{Australia}
 }
 \email{j.knibbe@uq.edu.au}
 \orcid{0000-0002-8844-8576}




\title{
Illusion Spaces in VR: The Interplay Between Size and Taper Angle Perception in Grasping}



\begin{abstract}

Leveraging the integration of visual and proprioceptive cues, research has uncovered various perception thresholds in VR that can be exploited to support haptic feedback for grasping. While previous studies have explored individual dimensions, such as size, the combined effect of multiple geometric properties on perceptual illusions remains poorly understood. We present a two-alternative forced choice study investigating the perceptual interplay between object size and taper angle. We introduce an illusion space model, providing detailed insights into how physical and virtual object configurations affect human perception. \add{Our insights reveal how, for example, as virtual sizes increase, users perceive that taper angles increase, and as virtual angles decrease, users overestimate sizes.}
We provide a mathematical model of the illusion space, \add{and an associated tool}, which can be used as a guide for the design of future VR haptic devices \add{and for proxy object selections.} 




\end{abstract}




\maketitle
\section{Introduction}

Accurate haptic feedback remains a grand challenge of virtual reality (VR). While visual and audio resolution continues toward realism, haptics remain relatively limited. 
One of the opportunities here is to leverage physical objects from the space around the user to provide haptic feedback for virtual objects in VR. By leveraging visuo-proprioceptive illusions, these objects do not have to be an exact match for one another. Understanding what virtual objects a physical object can represent remains a challenge. To date, research has considered the impact of individual geometric properties in isolation, and are yet to consider how different properties may impact one another and our perception.

This style of illusory, nearby-object-based haptic interaction is typically referred to as a passive haptics approach. These solutions benefit from being just-in-time, device-free, ad-hoc, and, (conceptually) scalable~\cite{clarence2021unscripted, clarence2024stacked,sharma2024graspui}. At the same time, they are constrained by perceptual illusion limits that require careful design and precise application to not adversely impact the user experience~\cite{gani2022impact}. 



\add{Much of the existing literature on passive haptics has explored object \textit{co-location} (considering \textit{where} the physical proxy object is, such that it may feel as though it is in the same location as the virtual object, e.g., through haptic retargeting~\cite{zenner2019estimating, clarence2021unscripted,clarence2024stacked,azmandian2016haptic}), rather than \textit{object similarity} (considering \textit{what} the object is, such that it may feel like the same in material and geometry) ~\cite{propping}.} This exploration of \textit{co-location} has revealed that objects can be approx. 15\textdegree{} around an arc to the left or right of the virtual object ~\cite{clarence2022investigating} and between 0.87x and 1.31x as far from the user~\cite{esmaeli_illusion}. 


\add{Explorations of \textit{similarity} have 
provided insights into the minimum and maximum size the proxy can represent~\cite{bergstrom2019resized}, and 
the perception of separate factors including sizes, angles and curves on the surface of the object~\cite{de2019different}. So far, these works have considered object properties in isolation. However, the properties' impacts on hand-pose, and joint forces and torques are not isolated -- size and shape together drive the finger spacing, orientations, and joint configurations of grasping, for example.} How object shapes and sizes interplay to impact our perceptual limits remains poorly understood.

We study the illusory limits and interplay of size and taper angle of physical proxies, \add{using real-time, virtual remapping techniques where users' fingers are retargeted whilst reaching for the objects}. 
We produced physical hexahedron objects for grasping, with 3 widths (3 cm, 6 cm, and 9 cm) and 3 taper angles (0°, 8°, and 16°, between the side face and the vertical direction). We conduct a study with 40 participants, grasping the physical objects with differing virtual counterparts (6 virtual sizes $\times$ 7 virtual angles for each physical object). 
By studying these object properties together, we reveal a complex, multi-dimensional \textit{illusion space} across both size and taper angle. 
For example, our results show how increasing the virtual size of an object will strongly increase the perceived angle of that object, while increasing the virtual angle makes the perceived size smaller. 
We provide a \addfinal{set of mathematical expressions} that 
describes a proxy object's haptic capacity and reveals how size and taper angle interplay with each other in creating illusions. We also discuss how these results may feed into mechanical, actuated, active haptics controller designs, to facilitate super-resolution haptic devices.

Our contributions:
\begin{enumerate}[nolistsep]
    \item  we study, reveal, and describe the interplay between size and taper angle when grasping objects in VR
    \item we provide visualisations of the multi-dimensional illusion space, to provide designers with guidelines on VR haptic perception
    \item we provide mathematical functions that enable designers to calculate whether a given physical object can represent an intended virtual object. 
    \item \add{We provide an online tool that allows designers to calculate physical-virtual object pairings within the illusion space.}
\end{enumerate}

\section{Related Work}

\add{When designing for advanced haptics in VR, researchers have to consider both co-location -- ensuring the haptic feedback is delivered in the correct location -- and object similarity -- ensuring the haptic feedback matches the material and geometric properties of the virtual object ~\cite{propping}. In this section, we begin by discussing the different approaches to increasing haptic fidelity, before presenting the current knowledge in illusory techniques and the associated perceptual limits.}

\subsection{\add{Providing Haptics in VR}}

Research has been exploring two avenues for achieving advanced haptics: (i) active haptics, and (ii) passive haptics. Some mechanically reflect the exact properties of the virtual objects (active controllers, e.g., \citet{choi2016wolverine}) while others use proxy objects and illusions to render them (passive approaches, e.g., \citet{bergstrom2019resized}). 

Much of the recent effort in active haptics has focused on handheld and wearable device designs. These can be portable and are most akin to the controllers we see being adopted commercially. Prominent examples here include wearable controllers for grasping rigid objects~\cite{choi2016wolverine,choi2018claw}, axisymmetric devices for grasping pseudo-cylindrical objects~\cite{10.1145/3472749.3474782}, and multiple degree-of-freedom controllers for grasping asymmetric objects~\cite{ulan2024}. To date, however, these controllers have remained specialist, cumbersome, and heavy. 

On the other hand, grounded and encounter-type devices provide just-in-time feedback, that is only available at the exact moment the user needs it and can leave them otherwise hands-free. Examples of such devices include the commercially available Touch\footnote{3D Systems, USA - formerly Phantom from Sensable Technologies} and Omega\footnote{Force Dimension, Switzerland} devices, alongside research prototypes such as inFORM~\citet{follmer2013inform}, ShapeShift~\cite{siu2018shapeshift}, and REACH+~\cite{gonzalez2020reach+}. While high-resolution, inFORM, for example, features 30$\times$30 individually controllable haptic pins~\cite{follmer2013inform}. This resolution highlights how these devices are often mechanically complex, bulky, expensive, and necessitate a step towards infrastructuring that makes them impractical for everyday use.

As a result of high-resolution active controllers remaining complex, specialist, and bespoke, research has also been exploring opportunities for leveraging \add{passive physical objects and haptic illusions, both to enable the use of physical objects in the user's reach for feedback and to extend the capacity of haptic controllers. 
Using passive haptic techniques has proven effective across a range of applications, including  for objects in the kitchen~\cite{hoffman1998physically}, for perception of ledges on the floor~\cite{insko2001passive}, to experience spiders~\cite{carlin1997virtual}, and to better leverage cups in the environment~\cite{simeone2015substitutional}. Furthermore, passive objects with simply designed geometry features can display various shapes with edges, curves, and surfaces by altering the visual presentation and taking advantage of VR visual feedback~\cite{ban2012modifying, ban2014displaying,yang2018vr}. }


\add{By more carefully considering and tailoring the physical props, however, we might be able to push the ability of the passive haptics further. By designing 3D printed materials and structures, for example, different textures can be simulated in VR~\cite{feick2023metareality, degraen2019enhancing}. As for the geometry properties, a toolkit named Haptwist was developed to combine passive proxies in a flexible way to represent more complex virtual objects~\cite{zhu2019haptwist}.} 







\add{Furthermore, there have been new methods developed in attempt to further push capabilities and to understand better users' perception in VR. To figure limits of hand redirection techniques, gaze detection and electroencephalogram (EEG) have been applied in some studies~\cite{feick2024predicting} instead of merely psychomotor metrics. Research has also explored the possibility of rendering haptics for complex, arbitrary shapes with visual feedback~\cite{zhao2018functional}.}




\add{These techniques seek to obscure the visuo-proprioceptive mismatches to convince the user they are interacting with the virtual object when, in fact, they are interacting with \textit{something else}, potentially \textit{somewhere else}, creating visuo-haptic illusions. When illusions are applied within perceptual limits, users are considered unlikely to notice their occurrence, which results in another type of haptic techniques using illusions such as retargeting~\cite{azmandian2016haptic} and resizing~\cite{bergstrom2019resized}. These techniques expand the ability of passive objects in VR haptics.}




\subsection{Illusions in Haptic Feedback}
 \begin{table*}
  \caption{\addfinal{Haptic illusion limits found by key previous studies on perception of size and angle}}
  \label{tab:previousstudy}

  \begin{tabular}{ll}
  
    \toprule
    \textbf{Study}&\textbf{Results}\\
    \hline
    \citet{barbagli2006haptic}&Force-direction threshold 31.9\textdegree{}\\
    \hline
    \citet{hajnal2011perceptual}&Overestimated the slope with fingers by around 6\textdegree{}\\
    \hline
    \citet{auda2021enabling}&Virtual size mismatch up to 50\% larger than the physical prop\\
    \hline
    \citet{bergstrom2019resized}&Physical size: 3 cm; virtual thresholds: 2.7 cm and 4.4 cm.\\ 
    &Physical size: 6 cm; virtual thresholds: 5.4 cm and 7.32 cm.\\
    &Physical size: 9 cm; virtual thresholds: 7.02 cm and 9.18 cm.\\
    \hline
    \citet{8816164}\footnotemark[3]& Physical width: 4 cm; virtual thresholds: 3.77 cm and 4.55 cm\\
    &Physical angle: 10\textdegree{}; virtual thresholds: 5.62\textdegree{} and 14.56\textdegree{}\\
    &Physical curvature: 33 m$^{-1}$; virtual thresholds: 11 m$^{-1}$ and 47.08 m$^{-1}$\\
    \bottomrule
\end{tabular}
\end{table*}
\add{Due to the influence and application potentials of illusions in VR haptics, numerous studies have concentrated on exploring how illusions affect users' perceptions. This illusory work has one of two focuses: \textit{co-location} (\textit{where} the object is), or \textit{similarity} (\textit{what} the object is)~\cite{propping}. }

\add{Illusions of \textit{co-location} have been heavily influenced by redirected touching~\cite{kohli2010redirected}, haptic retargeting~\cite{azmandian2016haptic} and haptic remapping~\cite{lohse2019leveraging}.} These illusions seek to guide the user's physical hand towards a proxy object that is spatially decoupled from its virtual counterpart. Examples have demonstrated redirect controller buttons~\cite{zenner2019estimating,gonzalez2019investigating}, enabled users to grab objects placed around them~\cite{clarence2022investigating}, and attempted to retarget random, unscripted reaches~\cite{clarence2021unscripted}. From work on these illusions, we have come to understand the spatial \textit{haptic coverage} of a physical object -- the area within which it can provide haptic feedback for virtual objects~\cite{clarence2024stacked}.

Illusions of \add{\textit{similarity}} explore the extent to which one physical object can feel like another. 
These illusions aim to convince users they are interacting with an object with one property (for example, a heavy hammer), while they in fact interact with an object with a different property or, at least, with an object with a different magnitude of that property (e.g., a lightweight bottle).
For example, illusions have been used to simulate factors in interaction such as geometry ~\cite{turchet2010influence}, force feedback ~\cite{lecuyer2000pseudo}, weight ~\cite{maehigashi2021virtual}, stiffness ~\cite{sanz2013elastic} and texture ~\cite{brahimaj2023cross}.
 To determine the limits of illusions in object perception, many studies have worked toward estimating the thresholds within which any difference between physical and virtual objects 
 cannot be reliably noticed. These limits now span a range of object properties (see Table \ref{tab:previousstudy}). For example, \citet{barbagli2006haptic} estimated the force-direction discrimination thresholds to be 31.9\textdegree{} when participants were given force-feedback with mismatched haptic and visual stimuli. \citet{hajnal2011perceptual} found users typically overestimated object slopes by around 6\textdegree{} without visual feedback.
 \citet{auda2021enabling} found that participants do not perceive the size mismatch for virtual objects up to 50\% larger than the physical prop. They also found that different displacement had no influence on the results. \citet{bergstrom2019resized} estimated the thresholds of resizing virtual objects from physical objects. They found that size perception thresholds varied when the sizes of the physical objects changed, which showed the thresholds could be influenced by different conditions. The 3 cm wide cuboid could represent virtual ones from 2.7 cm to 4.4 cm, while the 9 cm cuboid was only able to represent virtual ones from 7 cm to 9.2 cm. Similarly, \citet{8816164} estimated the thresholds of discrimination of sizes, face orientations and curvatures in VR separately. The results show Just Noticeable Difference (JND) values of 5.75\%, 43.8\%, and 66.66\% of the physical objects for the width, orientation of the prism faces, and curvature of ellipsoid, respectively. 

\footnotetext[3]{The upscaling thresholds are derived theoretical values from their study results.}

 \add{Importantly, across these works, reported perception thresholds differ, even when similar factors are being considered (for example \citet{bergstrom2019resized} and \citet{de2019different} report different size perception thresholds). This may result from a range of factors, including object properties, grasp specifics, interaction specifics, and more. Following this, Feick et al. specifically examined the impact of different factors on perceptual thresholds~\cite{feick2022designing}. They considered grasp type, movement trajectory, and object weight, but found no interplay. Crucially, their focus here was not on the perception of the grasp itself, but rather on the perception of a reaching movement. That is, they examined the impact of object properties on co-location, rather than similarity. We expect these factors (specifically grasp type and weight) to have a larger impact on similarity as they require alterations of grasp mechanics and joint forces~\cite{Stival2019}.} 
 
 \add{This suggestion is further supported by \citet{park2023visuo}, who found an interplay between the primary moment and product of inertia (MOI and POI, which can be roughly interpreted as differences in length -- MOI -- and shape and asymmetry -- POI). While participants waved the objects around during the study, the factors under discrimination were the geometric properties (thus, exploring object similarity). They report that participants were very sensitive to objects with a low MOI, regardless of the POI, but that haptic discrimination for a large POI degrades rapidly. POI and MOI are both object offset factors (i.e., they occur outside and away from the grasp). We seek to add further understanding to interplays in object similarities, by examining object properties within the grasp -- object size and shape.}


 \add{Varying object sizes directly impacts the joint orientations and torque forces experienced in the hand (primarily at the base of the fingers). Varying the shape of the object further alters the hand's pose, especially at the phalanges nearer the fingertips. It can also alter the direction of slippage at the fingertips, the direction of tangential forces applied by the fingers, and the cone of friction ~\cite{zheng2022human}. As a result of this combined impact on the biomechanics of the hand, interplays between factors must be examined directly and cannot be inferred through combinations of studies on individual factors. In order to understand these interplays, then, we study multi-dimensional estimation of thresholds.}

\section{Methods}

\begin{figure*}
  \includegraphics[width=0.75\textwidth]{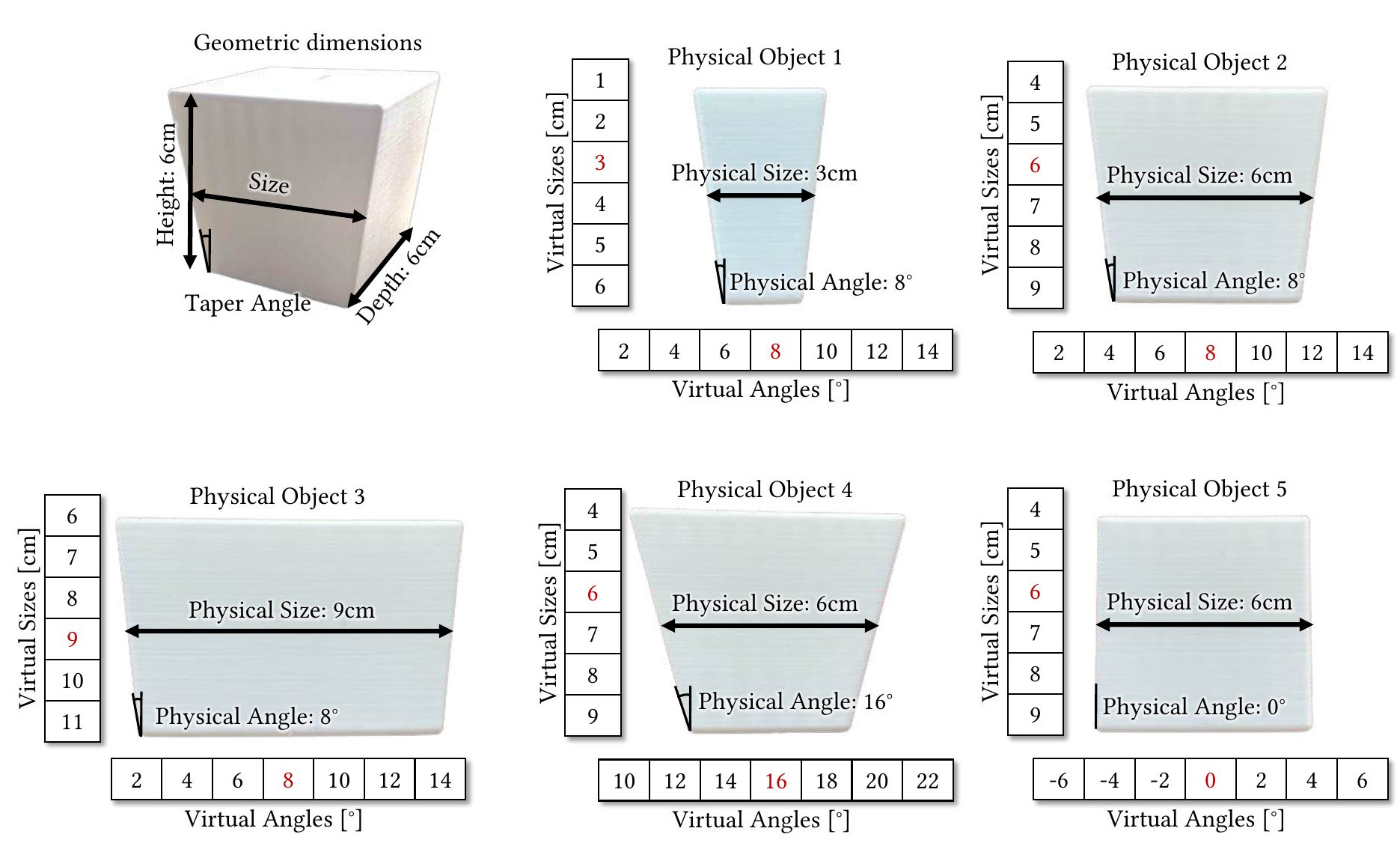}
  \caption{Physical objects for grasping. The dimensions of the objects are introduced on the top-left figure, and the following five figures are the physical objects in our experiment.}
  \Description{The definition of size, face orientation, height and depth of the objects.}
  \label{fig:cubephoto}
\end{figure*}

We conduct a study to model the perceptual thresholds for grasping physical objects as proxies for virtual objects in VR. Specifically, moving beyond the existing literature~\cite{bergstrom2019resized}, we consider the interplay between object size and shape (in this case, the taper angle of the surfaces being grasped). 

Where previous work has sought to provide psychomotor functions revealing upper and lower perceptual bounds (i.e., the points beyond which the user is likely to perceive the difference between the virtual and physical objects, thus harming the user experience~\cite{gani2022impact}), by considering multiple object properties together, we instead aim to provide a multi-dimensional illusion space which represents the perception thresholds with physical properties and the virtual ones considered simultaneously. This illusion space should serve to provide a richer insight into the range of virtual objects that a given physical proxy object can represent. 

Similar to prior work \addfinal{(.e.g,~\cite{bergstrom2019resized})}, we adopt a two-alternative forced choice (2AFC) paradigm
\addfinal{~\cite{KINGDOM2016149}} for our study. Under 2AFC, a user is forced to choose between two responses (e.g., smaller vs larger, or less tilted vs more tilted) even when they cannot tell the difference and their preferred choice would likely be 'the same'. In this scenario, the 2AFC paradigm assumes that they will select a response at random and become increasingly likely to select the correct response as they become more perceptually-certain. This technique enables the experimenter to identify a point-of-subject-equality (PSE), where the participants' choices are 50-50 and, thus, they perceive the different stimuli to be the same. Research then adopts 25\% and 75\% thresholds as the perceptual limits (see, e.g., ~\cite{bergstrom2019resized,de2019different,park2023visuo}), beyond which the participant is considered to be sufficiently certain in their response.

The GRASP taxonomy proposed by \citet{feix2015grasp} identifies 33 different grasp types. For our study, we chose to focus on the grasp involving only the index finger and thumb of the right hand \add{(Palmar Pinch, leveraging pad opposition)} due to its frequent use in tool operation and its suitability for handling objects weighing between 0 and 500 grams, with dimensions of 0 to 15 cm. This ``pinch-grasp'' is especially beneficial as it remains unaffected by the change in width from top to bottom in our tapered objects, ensuring consistent size perception.

\subsection{Experimental Conditions}

We produced five physical cubes of different sizes and taper angles, as shown in Fig.~\ref{fig:cubephoto}. 
\add{We chose the same sized cubes used by \citet{bergstrom2019resized} -- 3cm, 6cm, and 9cm. In their paper, these cubes had a taper angle of 0\textdegree{}. To study the interplay between size and taper angle (as a simple operationalisation of shape), we introduced two further angles  -- 8\textdegree{} and 16\textdegree{}. As 2AFC studies require repeated measures, and in order to make our study tractable in size, we} chose the following combinations for the physical objects: 3cm-8\textdegree{}, 6cm-8\textdegree{}, 9cm-8\textdegree{}, 6cm-16\textdegree{}, and 6cm-0\textdegree{}.


The sizes defined for the cubes are the widths halfway up the cube. 
There are six different virtual sizes for each physical object size (distributed around the physical size, as shown in Fig.~\ref{fig:cubephoto}). The step of virtual sizes is 1cm. The maximum width of grasping with the thumb and the index finger is 12.42cm (SD=1.40cm) on average for an adult human~\cite{garrett1971adult}. Therefore the size of our largest virtual cube is set to be 11cm and the size of our largest physical cube is set to be 9cm which is about 80\% of the lower end of the standard deviation. \add{The physical and virtual settings of sizes are the same as in prior work~\cite{bergstrom2019resized}.} The height and depth for all physical and virtual cubes are 6cm. 

The taper angles of the cubes are angular-degree off the upright of the bottom corner. This results in a inverted trapezoid object -- a shape that can be held steadily by the participants \add{with a more stable cone of friction~\cite{zheng2022human} and less potential for slippage. This constraint was selected to reduce confounding factors in the study.}
For the virtual cubes, there are seven different angles with step of $2^\circ$ for each physical cube. This step was determined through pilot testing. For the cube with $0^\circ$ face orientation, there are virtual inverted trapezoid regular trapezoid objects. The negative figures of angles in Fig.~\ref{fig:cubephoto} mean the shape of virtual cube sections are regular trapezoid.


\begin{figure*}
  \includegraphics[width=0.85\textwidth]{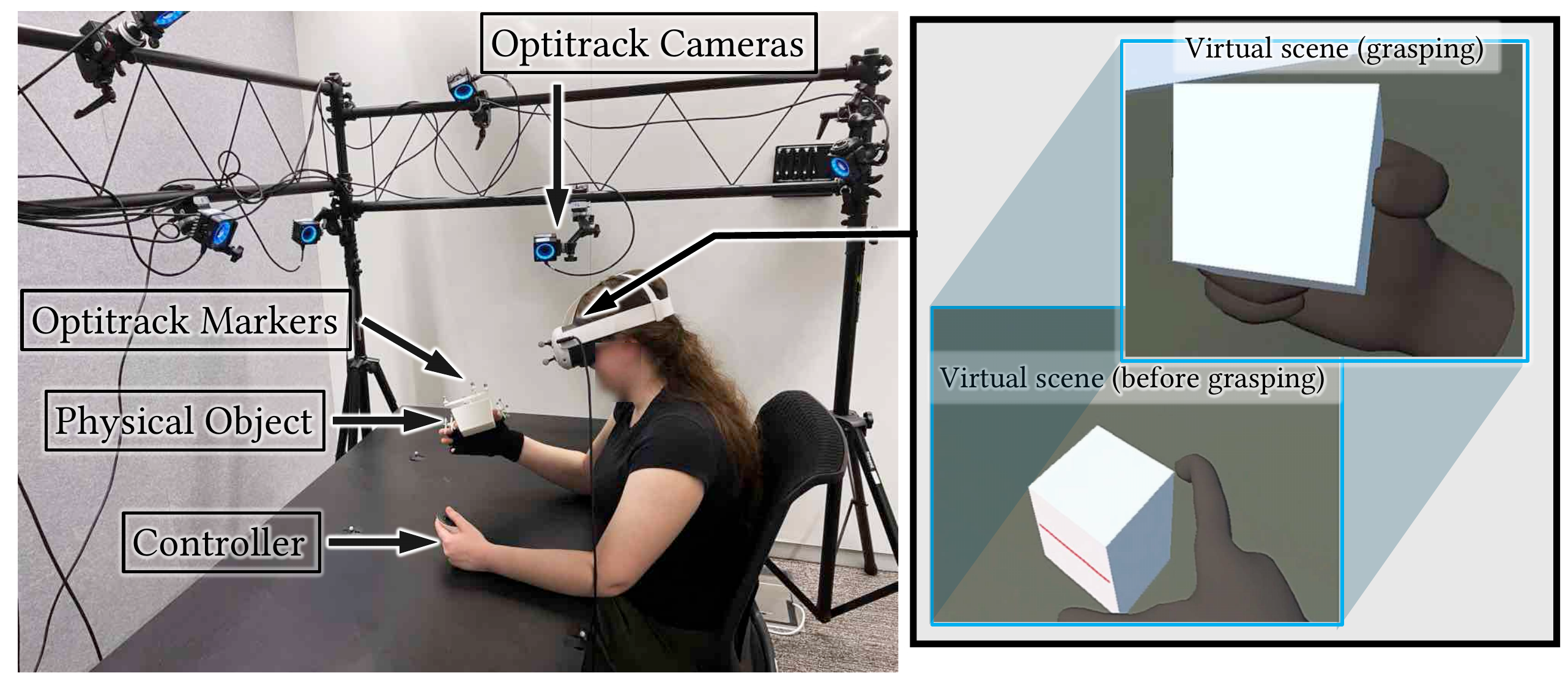}
  \caption{\add{Figures showcasing the experimental setup. Left: the study apparatus, including the participant's seating position, Optitrack setup for object tracking, and physical object markers. Right: the virtual view before grasping (highlighting the red center line) and during grasping.}}
  \Description{The definition of size, face orientation, height and depth of the objects.}
  \label{fig:setting}
\end{figure*}

The experiment consisted of lifting the five physical objects, each of them used as a proxy for virtual cubes across six distinct sizes and seven different taper angles. Three of the designed virtual cubes are infeasible (size of 1cm and taper angle of $10^\circ, 12^\circ$ and $14^\circ$ -- objects where the sides would cross themselves and so violate our height requirements).
Therefore, the total number of experimental conditions is 207: 5 physical objects $\times$ 6 virtual sizes $\times$ 7 virtual angles - 3 infeasible shapes.

\subsection{Apparatus}
All the physical objects were 3D printed with polylactic acid (PLA). Their weight was controlled to be roughly identical despite a range of geometries by adjusting the printing density ($67.98g\pm0.26g$). 

 We used a Meta Quest 3 HMD with the Optitrack motion capture system for tracking the hand motion and the cubes. Seven Prime 13W cameras were placed around the table and set at a tracking frequency of 240Hz (see Fig.~\ref{fig:setting}). We designed rigid tracking bodies of different shapes to incorporate the tracking markers which were 
 attached to the physical cubes and participants' index finger and thumb. These rigid bodies were uniform in weight and did not interfere with the participants' range of motion or object interactions (they were mounted above the objects to be grasped). The VR scene was built in Unity 2022 on a laptop PC (13th Gen Intel i9-13900HK 2.60GHz, 32.0GB RAM, NVIDIA GeForce RTX 4090 GPU, Windows 11 Pro). FinalIK by Rootmotion~\cite{finalIK} was used as inverse kinematic model for finger motions as a Unity asset.

\subsection{Hand Rendering}
We applied resized grasping techniques similar to those used in a previous study by \citet{bergstrom2019resized} to render the hand model. In the virtual scene, only the index finger and thumb were animated on the hand. The positions of the participant's hand, index finger and thumb were captured with motion capture system and streamed into Unity. The distance of the index finger and thumb to be rendered in the virtual scene $D_{v}$ was calculated based on the scaling ratio between the width of the virtual cube and the physical cube $S_{v}/S_{p}$, where $S_{v}$ is the size of the virtual cube and $S_{p}$ is the size of the physical cube. The virtual hand was resized by the method to have a distance $D_{v}$ between the virtual index finger and thumb:
\small
\begin{equation}
    D_{v} = D_{p}\times(S_{v}/S_{p})
\end{equation}
\normalsize

where $D_{p}$ is the distance between index finger and thumb in real world.

The inverse kinematic model was then used to render the motion of the finger joints based the new index finger tip and thumb tip positions. While previous work applied the finger distance changes during the reach motion~\cite{bergstrom2019resized}, \add{we applied the real-time scaling algorithm to the hand across the whole grasping from the beginning of each task to avoid participants noticing finger distance changing when approaching the objects.}

\subsection{Experimental Procedure}

\begin{figure*}
  \includegraphics[width=0.95\textwidth]{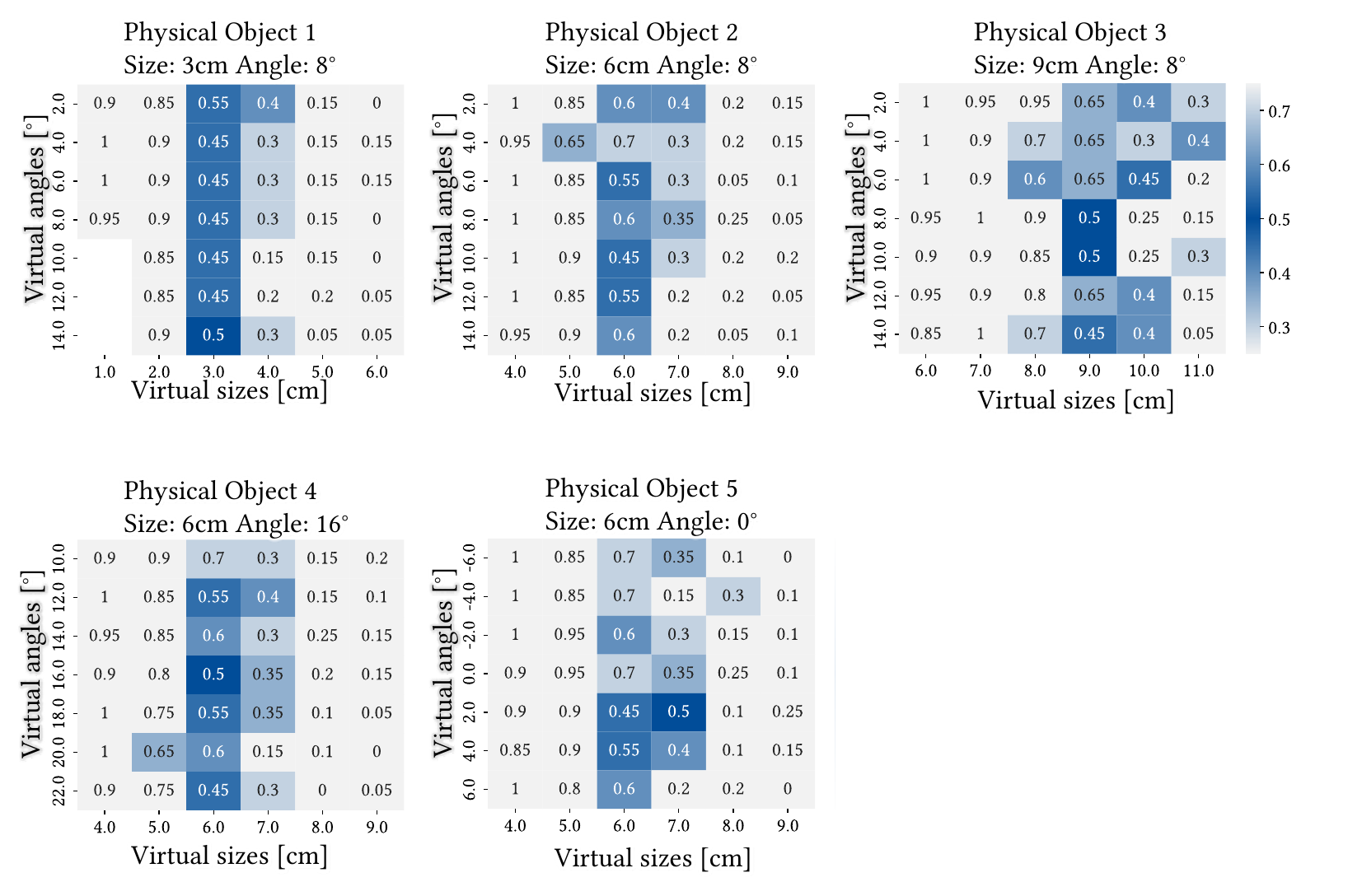}
  \caption{Proportion of choosing ''virtual smaller''. \add{Blue areas show instances configurations in which the illusion holds, where grey areas indicate cube pairings where the illusions break.} 
  }
  \Description{}
  \label{fig:sizeheatmap}
\end{figure*}

Upon recommendation from our local ethics committee, the participants were limited to being in VR for one hour. Considering the workload of grasping 207 cubes and responding to questions, we divide the task between two participants. Therefore, for each pair of participants, the 207 combinations of grasping objects were randomised and split into two parts (103 and 104 grasps). 

Participants used their right hands to grasp the physical objects while looking at the virtual objects in the VR headset. 
Similarly to previous studies~\cite{caparos2009interacting}, we asked questions that  were rendered in the virtual scene: ``Is the virtual cube smaller or larger than the physical cube?'' and ``Is the virtual cube less tilted or more tilted than the virtual cube?''. Participants responded to these questions with a controller held in their left hand. 

To perform the task, participants were asked to reach and grasp the object on the red lines on the middle height level of the two tilted faces - this ensured the participants grasped on the sides with the taper angle and the widths were controlled here. The participants then lifted the object and answered the questions while holding the object. \add{The two questions were presented at the same time and the participants could answer them in any order.} After answering both questions they could put down the object and move their right hands back to the edge of the table, at which point the virtual scene would reset. During this time, the researchers replaced the physical objects even when the physical objects didn't change. The participants could rotate the object to observe it but they were not permitted to move their fingers along the surfaces of the object \add{as this could have served to reveal more geometric information about the object in question.} 

Prior to beginning, we asked each participant to perform three example object grasps (with physical and virtual properties: 6cm-16\textdegree{} and 6cm-0\textdegree{}, 9cm-8\textdegree{} and 3cm-0\textdegree{}, 9cm-8\textdegree{} and 9cm-16\textdegree{}, showcasing both size and angle incongruences) for training. The experiment commenced right after. The participants had two breaks during the experiments. The procedure took 50 min on average. \add{During the study, the experimenter monitored the participants, to ensure the tasks were completed correctly (i.e., the participants lifted the objects from the centre line). \addfinal{In case of failure, the experimenter would repeat failed trials later during the study. However, no noticeable errors were made by participants. A video figure of the experimental setup and procedure can be found in the supplementary materials.}}


\subsection{Participants}
We recruited 40 right-handed participants contributing to 20 data points for each grasping combination. \add{The recruitment information was posted on a public university website.} \addfinal{Among the participants, 21 of them identified themselves as females and 19 as males.} Their ages ranged from 18 to 38 and the average age was 24.6 (SD=4.1). Their average thumb-index finger span was 16.6cm (SD=2.0). \addfinal{There were 14 participants who claimed they had no prior experience in VR, while 25 claimed some experience and 1 claimed extensive experience.} The study was approved by the Ethics Board of the university. Each participant received a gift card \add{with the value of 20 USD} as compensation for participating in the experiment.

\section{Results}

\begin{figure*}
  \includegraphics[width=0.95\textwidth]{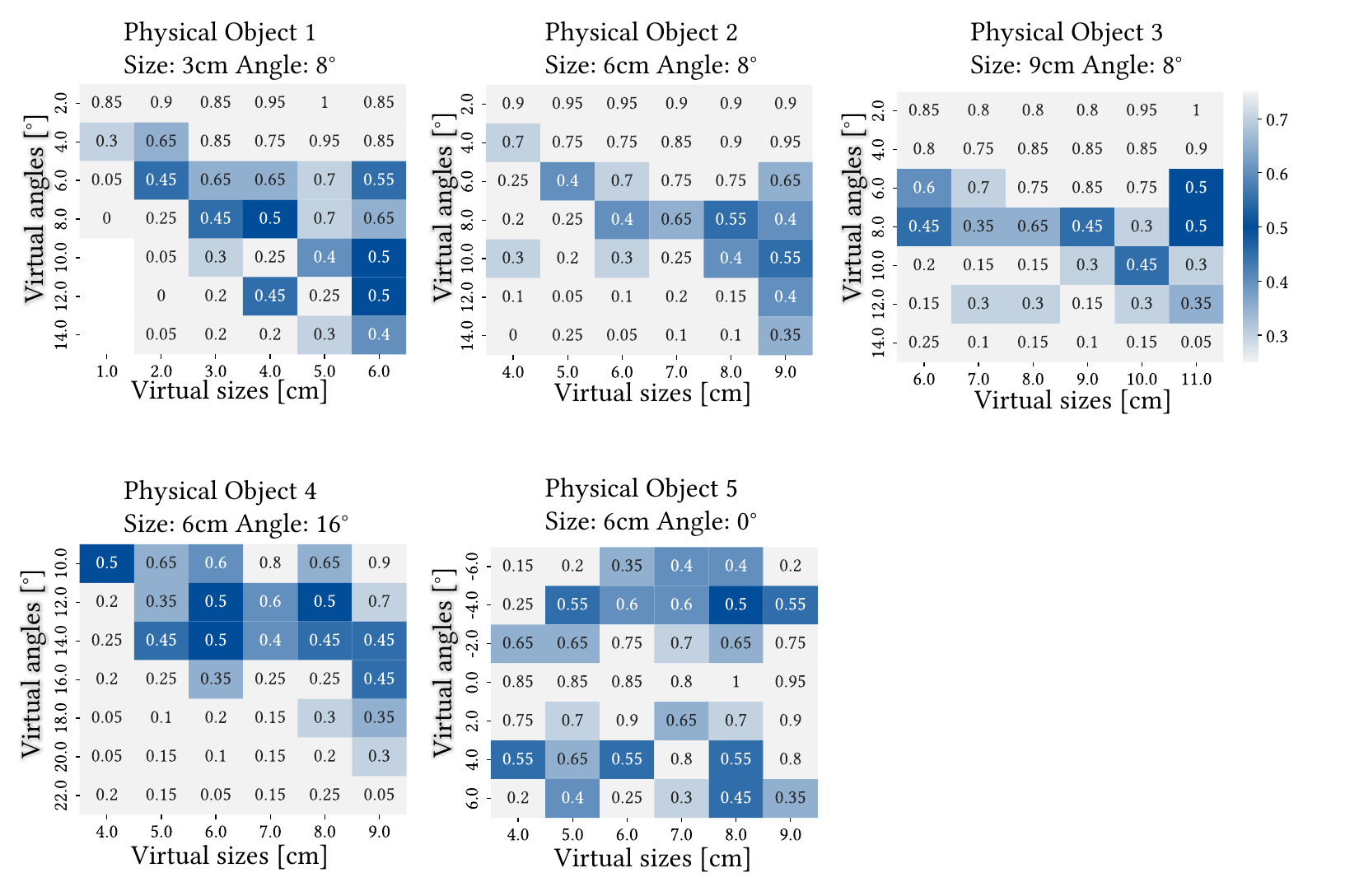}
  \caption{Proportion of choosing ''virtual less tilted''. \add{Blue areas show instances configurations in which the illusion holds, where grey areas indicate cube pairings where the illusions break.}
  }
  \Description{}
  \label{fig:angleheatmap}
\end{figure*}
To begin, we calculated the proportions of responses to the 2AFC questions. 
If less then 25\% or more than 75\% of participants select ''virtual smaller'' or ''virtual less tilted'', then there is a 75\% probability for them to notice differences between physical (haptic) and virtual (visual) stimulus. 
Following this, illusions are considered effective when the proportions of responses are between 25\% and 75\%. This forms the illusion space. The proportions of participants selecting ''virtual smaller'' are shown in Fig.\ref{fig:sizeheatmap}. In this heatmap, the coloured cells show the illusion space where participants tend to believe they are grasping objects with the same properties in virtual and physical scenes, while the grey cells show when the illusion breaks. As the virtual sizes increase, the proportion of selecting ''virtual smaller'' drops. The proportions of answering ''virtual less tilted'' is shown in Fig.~\ref{fig:angleheatmap}. When the values of the virtual angles are smaller, the proportion of selecting ''virtual less tilted'' increases. According to the results, given the object step sizes we chose (1 cm per size step and 2\textdegree{} per taper angle step), the illusion space for angle perception is bigger than for size. Both illustrations (Fig.\ref{fig:sizeheatmap} and Fig.\ref{fig:angleheatmap})  show the perceptual interplay between size and taper angle, with non-uniform distributions of proportions across the factors. Specially, the virtual size seems to have a significant influence on the angle perception. As shown in Fig. \ref{fig:angleheatmap}, the perceived illusion of angle expands greatly as virtual object sizes increase. Interestingly,  participants sometimes held objects with identical virtual and physical angles but they wouldn't perceive this was the case when the virtual size was smaller than the physical objects.

\subsection{Perception Thresholds with Single Property Incongruence of Size and Taper Angle}

We first present the individual results for sizes and taper angles as single-dimensional analysis (either only the virtual size or only the virtual angle changes). 
The proportions of answers have been processed similarly to previous studies based on the answers of 2AFC questions \cite{bergstrom2019resized, de2019different}. For the size perception, the percentage of selecting ``virtual smaller'' is calculated for each different virtual size combination while the other property of the virtual objects, the taper angle, is congruent with that of the physical objects (i.e., the same as the physical objects). The same approach is applied to taper angle perception, with the virtual size being congruent with the physical objects' size. The data points are fitted to the sigmoid function:
\small
\begin{equation}
f(x)=\frac{1}{1+e^{ax+b}}
\label{eq:sigmoid}
\end{equation}
\normalsize
For the size perception, the point at which there is a 50\% chance of selecting ``virtual smaller'' is considered the Point of Subjective Equality (PSE); where the virtual object is estimated to be the same size as the physical object (e.g. the participants randomly select from 'smaller' or 'larger', as they perceive the objects to be the same size). \addfinal{The 25\% and 75\% points, which are used to describe the perception limits in similar psychological studies (e.g., ~\cite{bergstrom2019resized,de2019different,park2023visuo}),} are correspondingly selected as the upscaling thresholds (UT) and downscaling thresholds (DT) of perception, where the participants become able to accurately determine differences between the physical and virtual objects with 75\% certainty. 
The Just-Noticable Difference (JND) is defined as the distance between the upscaling or downscaling threshold and the PSE. We can also obtain the Weber fraction, reflecting how sensitive users are to changes in the sensory stimulus by calculating the ratio of JND to the reference physical size. The same approach is also applied to the taper angle perception results based on percentages of selecting ``virtual less tilted'' (except grasping tasks with the 6cm-0\textdegree{} physical object, which is analysed separately because of its unique taper orientation).

\begin{figure*}
  \includegraphics[width=0.95\textwidth]{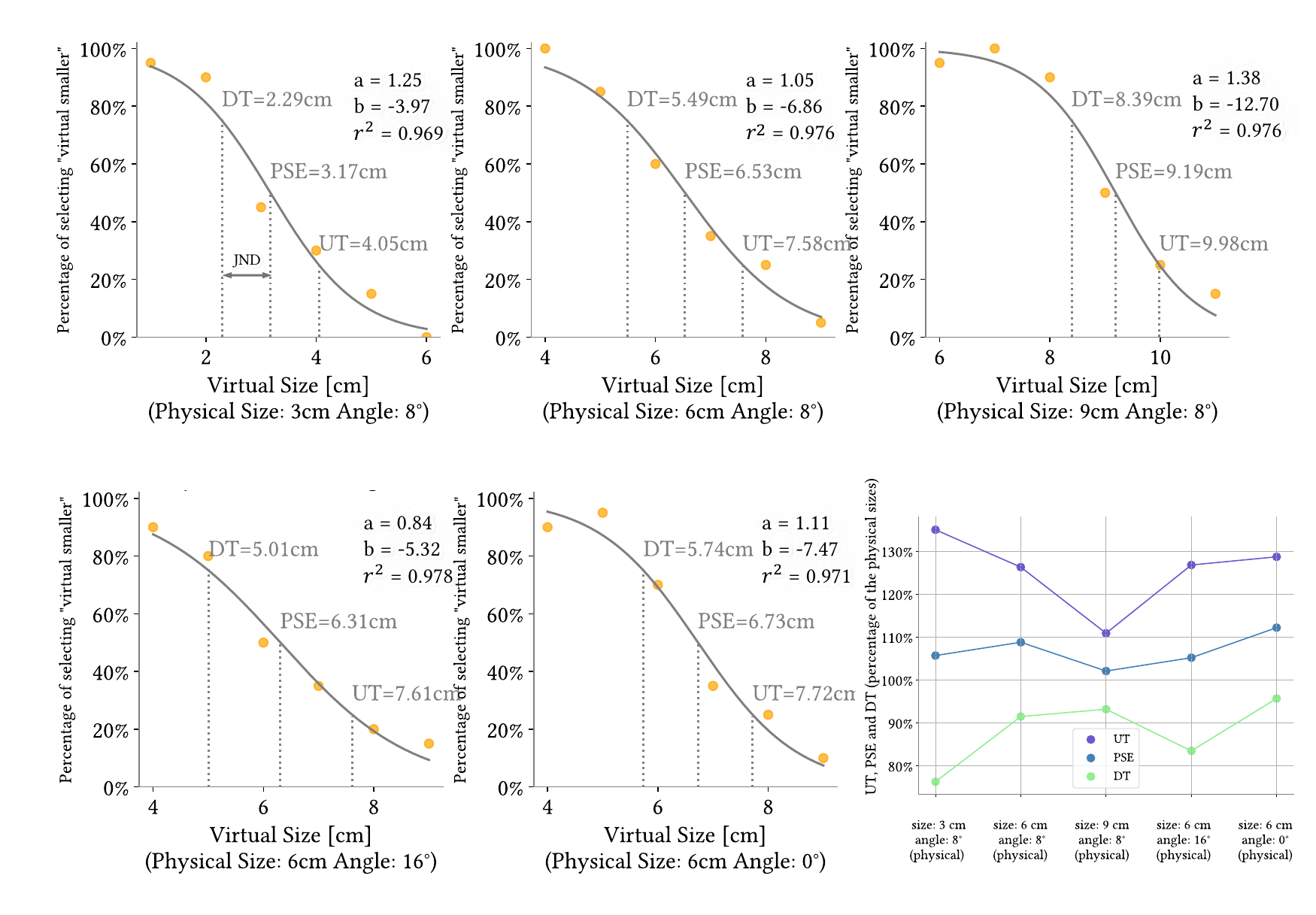}
  \caption{Size perception results with congruent virtual angles. Graphs show the point of subjective equality (PSE) and upscaling and downscaling thresholds (UT and DT, respectively). The sigmoid function coefficients of each curve are provided.}
  \Description{}
  \label{fig:size1}
\end{figure*}

The size perception results of the five physical objects are shown in Fig. \ref{fig:size1}. These figures demonstrate the range of virtual objects that each physical object can represent (through the space between the downscaling threshold -- DT -- and the upscaling threshold -- UT). 

\add{These results show that,} for the three 8\textdegree{} objects, the range of the illusion space (formed from 2*JND, \addfinal{expressed as a percentage}) gets smaller as the physical object's size increases. Across the three 6cm physical objects, as the taper angle increases, the PSE gets smaller towards the true size of the physical object, and the illusion spaces are also larger. 




\begin{figure*}
  \includegraphics[width=0.95\textwidth]{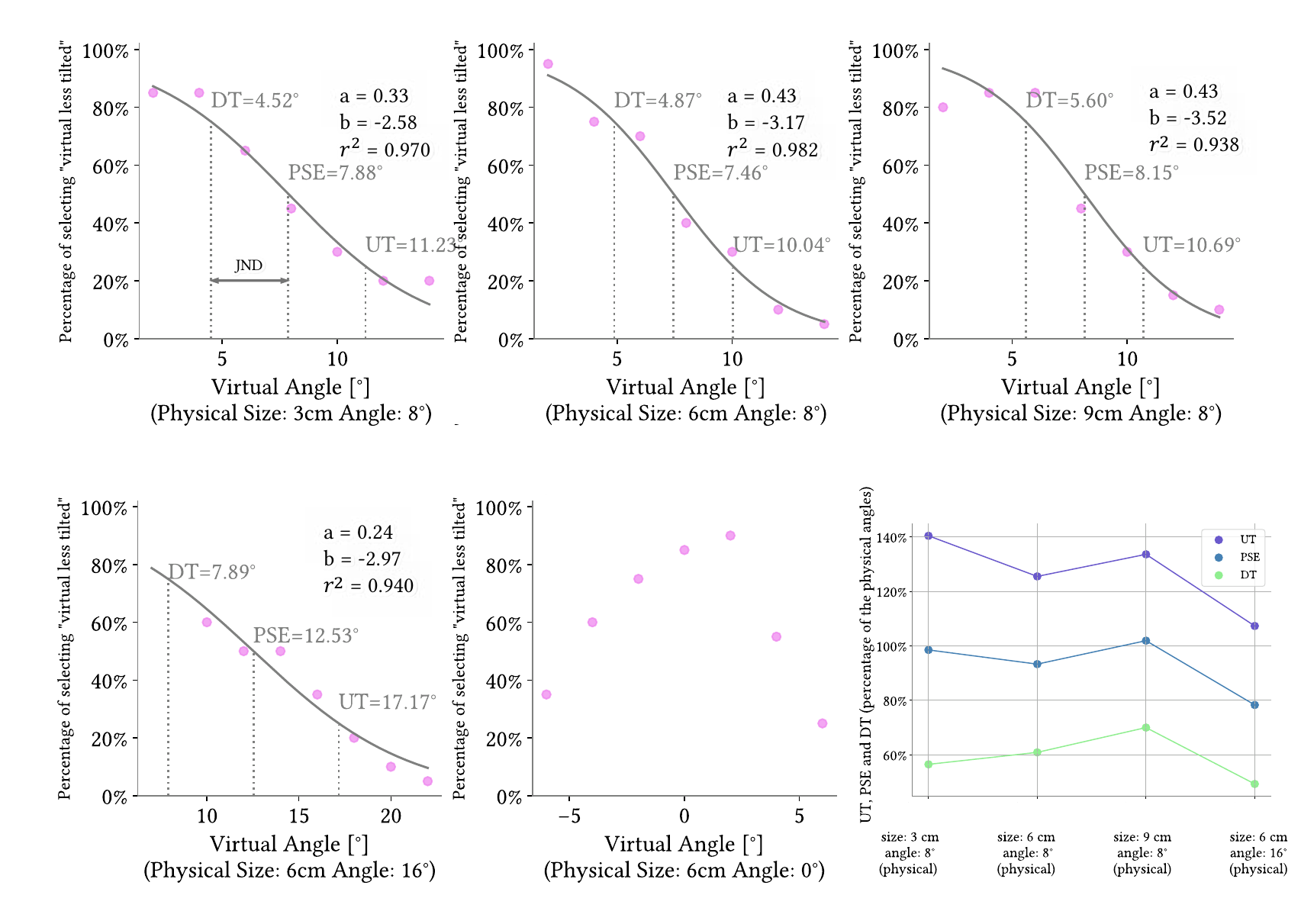}
  \caption{Angle perception results with congruent virtual sizes. Graphs show the point of subjective equality (PSE) and upscaling and downscaling thresholds (UT and DT, respectively). The sigmoid function coefficients of each curve are provided.}
  \Description{}
  \label{fig:angle1}
\end{figure*}

\delete{There are three physical objects with an 8\textdegree{} taper angle: 3cm-8\textdegree{}, 6cm-8\textdegree{}, and 9cm-8\textdegree{}. 
The fit of physical object 3cm-8\textdegree{} is plotted with $r^2=0.97$. The thresholds are 4.52\textdegree{} (56.5\%) and 11.23\textdegree{} (140.4\%). The PSE is 7.88\textdegree{} (98.5\%). The JND is 3.36\textdegree{} (Weber fraction 42.0\%). 

The fit of the 6cm-8\textdegree{} object is plotted with $r^2=0.98$. The detection thresholds are 4.87\textdegree{} (60.9\%) and 10.04\textdegree{} (125.5\%). The PSE is 7.46\textdegree{} (93.3\%) and the JND is 2.59\textdegree{} (Weber fraction 32.4\%).

The fit of physical object 3 is plotted with $r^2=0.94$. The detection thresholds are 5.60\textdegree{} (70.0\%) and 10.69\textdegree{} (133.6\%). The PSE is 8.15\textdegree{} (101.9\%) and the JND is 2.55\textdegree{} (Weber fraction 31.9\%).

The 6cm-16\textdegree{} object is plotted with a goodness of fit of $r^2=0.94$, although one of the thresholds estimated (7.89\textdegree{}) is out of the range of experiment settings (10\textdegree{}-22\textdegree{}). The estimated detection thresholds are 7.89\textdegree{} (49.3\%) and 17.17\textdegree{} (107.3\%). The PSE is 12.53 (78.3\%) and the JND is 4.64\textdegree{} (Weber fraction 29.0\%).}

\add{The results of angle perception are shown in Fig. \ref{fig:angle1}.} These results showcase how the just noticeable difference of taper angle decreases as object sizes get larger, but that absolute equality perception is the least accurate for mid-sized objects. For objects with larger taper angles, participants underestimate the angle of the physical object. These results also reveal that the physical object of 6cm-0\textdegree{} has different patterns from the others. As the virtual angle ranges of this object include both inverted and regular trapezoid shapes, any angle perceived to be non-zero is `more tilted' than the physical object. This leads to a different pattern of results for this object. 
However, this data requires a different analysis function and violates the underlying principles of the 2AFC method (when the object is perceived to be the same, it is not clear that the participants would be choosing between the answers completely at random). Because of the ambiguity of the 2AFC question, we cannot derive a sigmoid for this data. Thus, we choose not to analyse this data further.  

\subsection{Perception Thresholds with Both Size and Angle Incongruences}
Previously, the ability of a physical object to represent different virtual objects has been estimated with only the incongruence of a single property (e.g., size, taper angle, or curvature~\cite{bergstrom2019resized, park2023visuo, de2019different}) while all the other properties are congruent between the physical and virtual objects. Exploring how the detection thresholds change with multiple incongruent properties can expand the possibility for physical objects to represent virtual ones. In addition to the thresholds estimated for each physical object and each property, the patterns of how the thresholds change with the incongruence of the other property are demonstrated here. 


Fig. \ref{fig:size1} shows the size perception thresholds and PSEs when the virtual angles are congruent with the physical ones. When the angles are incongruent, e.g. the physical angle of 8\textdegree{} and the virtual angle of 10\textdegree{}, a fit can also be plotted and thresholds can be calculated using the same method. Therefore, for one physical object, when being presented with different virtual angles, the results of the perception of size are estimated. The same approach is also applied to angle perception results with size incongruences.

For all physical objects, the size perception results are fitted to a sigmoid function (\ref{eq:sigmoid}) and the upscaling threshold (UT), downscaling threshold (DT), PSE and JND are calculated.

The results of perception estimation of physical objects 1-5 are shown in Table \ref{tab:sizeinc} - Table \ref{tab:both1}.

\begin{table*}
  \caption{Size perception estimation with different virtual angles}
  \label{tab:sizeinc}
  \begin{tabular}{cllllllll}
    \toprule
    \textbf{Physical Object 1}&Virtual Angle [\textdegree{}]&2&4&6&8&10&12&14\\
    \textbf{size: 3cm, angle: 8$^\circ$}&UT [cm]&4.43&4.16&4.16&4.05&3.67&3.95&3.99\\
    &DT [cm]&2.36&2.30&2.30&2.29&2.23&2.12&2.40\\
    &PSE [cm]&3.39&3.23&3.23&3.17&2.95&3.04&3.20\\
    \hline
    \textbf{Physical Object 2}&Virtual Angle [\textdegree{}]&2&4&6&8&10&12&14\\
    \textbf{size: 6cm, angle: 8$^\circ$}&UT [cm]&7.73&7.71&7.07&7.58&7.37&7.04&6.88\\
    &DT [cm]&5.49&5.10&5.43&5.49&5.25&5.36&5.58\\
    &PSE [cm]&6.61&6.40&6.25&6.53&6.31&6.20&6.23\\
    \hline
    \textbf{Physical Object 3}&Virtual Angle [\textdegree{}]&2&4&6&8&10&12&14\\
    \textbf{size: 9cm, angle: 8$^\circ$}&UT [cm]&10.89&11.22&10.95&9.98&10.51&10.63&10.19\\
    &DT [cm]&8.74&8.05&7.99&8.39&8.02&8.40&7.92\\
    &PSE [cm]&9.82&9.63&9.47&9.19&9.27&9.51&9.05\\
    \hline
    \textbf{Physical Object 4}&Virtual Angle [\textdegree{}]&10&12&14&16&18&20&22\\
    \textbf{size: 6cm, angle: 16$^\circ$}&UT [cm]&7.54&7.50&7.66&7.61&7.25&6.86&6.89\\
    &DT [cm]&5.60&5.43&5.34&5.01&5.25&5.10&4.97\\
    &PSE [cm]&6.57&6.47&6.50&6.31&6.25&5.98&5.93\\
    \hline
    \textbf{Physical Object 5}&Virtual Angle [\textdegree{}]&-6&-4&-2&0&2&4&6\\
    \textbf{size: 6cm, angle: 0$^\circ$}&UT [cm]&7.31&7.33&7.25&7.72&7.83&7.53&7.09\\
    &DT [cm]&5.76&5.53&5.64&5.74&5.20&5.33&5.37\\
    &PSE [cm]&6.54&6.43&6.45&6.73&6.52&6.43&6.23\\
    \bottomrule
\end{tabular}
\end{table*}

\begin{table*}
  \caption{Angle perception estimation with different virtual sizes}
  \label{tab:both1}
  \begin{tabular}{cllllllll}
    \toprule
    \textbf{Physical Object 1}&Virtual Size [cm]&1&2&3&4&5&6&\\
    \textbf{size: 3cm, angle: 8$^\circ$}&UT [\textdegree{}]&4.22&7.62&11.23&12.69&12.79&17.86\\
    &DT [\textdegree{}]&2.48&3.46&4.52&4.29&6.51&4.31\\
    &PSE [\textdegree{}]&3.35&5.54&7.88&8.49&9.65&11.08\\
    \hline
    \textbf{Physical Object 2}&Virtual Size [cm]&4&5&6&7&8&9&\\
    \textbf{size: 6cm, angle: 8$^\circ$}&UT [\textdegree{}]&7.70&8.35&10.04&11.06&11.21&14.86\\
    &DT [\textdegree{}]&2.88&3.59&4.87&6.00&6.02&5.03\\
    &PSE [\textdegree{}]&5.29&5.97&7.46&8.53&8.62&9.94\\
    \hline
    \textbf{Physical Object 3}&Virtual Size [cm]&6&7&8&9&10&11&\\
    \textbf{size: 9cm, angle: 8$^\circ$}&UT [\textdegree{}]&10.61&10.39&11.49&10.69&11.77&11.13\\
    &DT [\textdegree{}]&4.01&3.78&5.34&5.60&5.02&4.87\\
    &PSE [\textdegree{}]&7.31&7.08&8.42&8.15&8.39&8.00\\
    \hline
    \textbf{Physical Object 4}&Virtual Size [cm]&4&5&6&7&8&9&\\
    \textbf{size: 6cm, angle: 16$^\circ$}&UT [\textdegree{}]&13.61&16.27&17.17&16.55&18.97&19.1\\
    &DT [\textdegree{}]&3.25&6.70&7.89&9.93&5.85&11.16\\
    &PSE [\textdegree{}]&8.43&11.48&12.53&13.24&12.41&15.13\\
    \bottomrule
\end{tabular}
\end{table*}

\section{Derived Analysis}
The results from our modelling demonstrate that the ability of physical objects to represent different virtual objects varies when either physical or virtual properties change. Our results reveal a non-uniform interplay between object sizes and taper angles and, as such, vary from those seen in previous studies of size and taper angle perception individually~\cite{bergstrom2019resized, de2019different}. We begin by discussing how we build illusion spaces from our results and how those spaces compare to results seen in prior work. We discuss the patterns we see in the perception thresholds and points of subjective equality and how these change with the interplay between size and taper angle. \add{At each step of the analysis the sizes and taper angles are coupled and thus we can add dimensions gradually on the illusion spaces model to reveal the coupling patterns.} Based on our results, we derive a broader, multi-dimensional illusion space across size and taper angle.

\subsection{Single-dimensional Illusion Space}

For each physical object without virtual angle incongruence (i.e., where the virtual angle is the same as the physical angle), a single-dimensional illusion space is formed from the upscaling threshold (UT) and downscaling threshold (DT) of size perception (these are the points beyond which participants can reliably distinguish between physical and virtual object' sizes. The distance between these thresholds is \addfinal{2*JND (expressed as a percentage)} which is also \addfinal{hereby defined as} the range of the illusion space. Within the confines of this single-dimensional illusion space, the size differences between the physical and the virtual objects are not noticeable. We use the same approach to construct illusion spaces for the taper angles, where the virtual sizes are congruent with the physical. 

Fig. \ref{fig:size1} shows the size thresholds of perception for our five physical objects. These results reveal differences with those seen in the prior work of 
\citet{bergstrom2019resized}, where the perception thresholds for physical objects of 3cm, 6cm and 9cm were 89\%-146\% (JND: 21.5\%), 90\%-122\% (JND: 16.0\%) and 78\%-102\% (JND: 12.0\%), respectively.  For our group of 3cm, 6cm, and 9cm physical objects with a taper angle of 8\textdegree{}, the thresholds are 76.3\% -135.0\% (JND: 29.4\%), 91.5\%-126.3\% (JND: 17.4\%) and 93.2\%-110.9\% (JND: 8.85\%). While the thresholds vary, the corresponding JNDs are similar for each size. This shows that the ranges of the illusion spaces (from JNDs) are similar, but that the consistent 8\textdegree{} taper angle makes smaller objects feel smaller and larger objects feel larger. 
The size perception results of our 6cm-0\textdegree{} object (95.7\%-128.7\% - JND: 16.5\%) are very similar to \citet{bergstrom2019resized}'s. This helps to confirm that the differences seen in our results stem from the impact of the taper angle. 

In another prior study on size perception, \citet{de2019different} found a JND for a 4cm object of 9.75\%. This is quite different to the results seen in our study and in \citet{bergstrom2019resized}'s study (based on the interpolation of our 3cm and 6cm results, as neither our study nor the Resized Grasping study considered 4cm objects per se). These differences could exist for a range of reasons, and likely highlight the potential interplay of some other property or factor. For example, research has shown how surface textures and weight of physical objects can affect the pressure required for grasping~\cite{maehigashi2021virtual}, and this would likely impact size perception. 


The JNDs for angle perception in our study for the physical objects 3cm-8\textdegree{}, 6cm-8\textdegree{}, 9cm-8\textdegree{} and 6cm-16\textdegree{} are 42.0\%, 32.4\%, 31.9\% and 29.0\% respectively. In a prior study, de Tinguy et al.~\cite{de2019different} estimated the JND for a 4cm physical object with 10\textdegree{} taper angle was 44.7\% which is similar to the 42.0\% JND of our 3cm physical object with 8\textdegree{} taper angle.

\subsection{Two-dimensional Illusion Space: both virtual size and angle incongruent}
\begin{figure*}
  \includegraphics[width=0.95\textwidth]{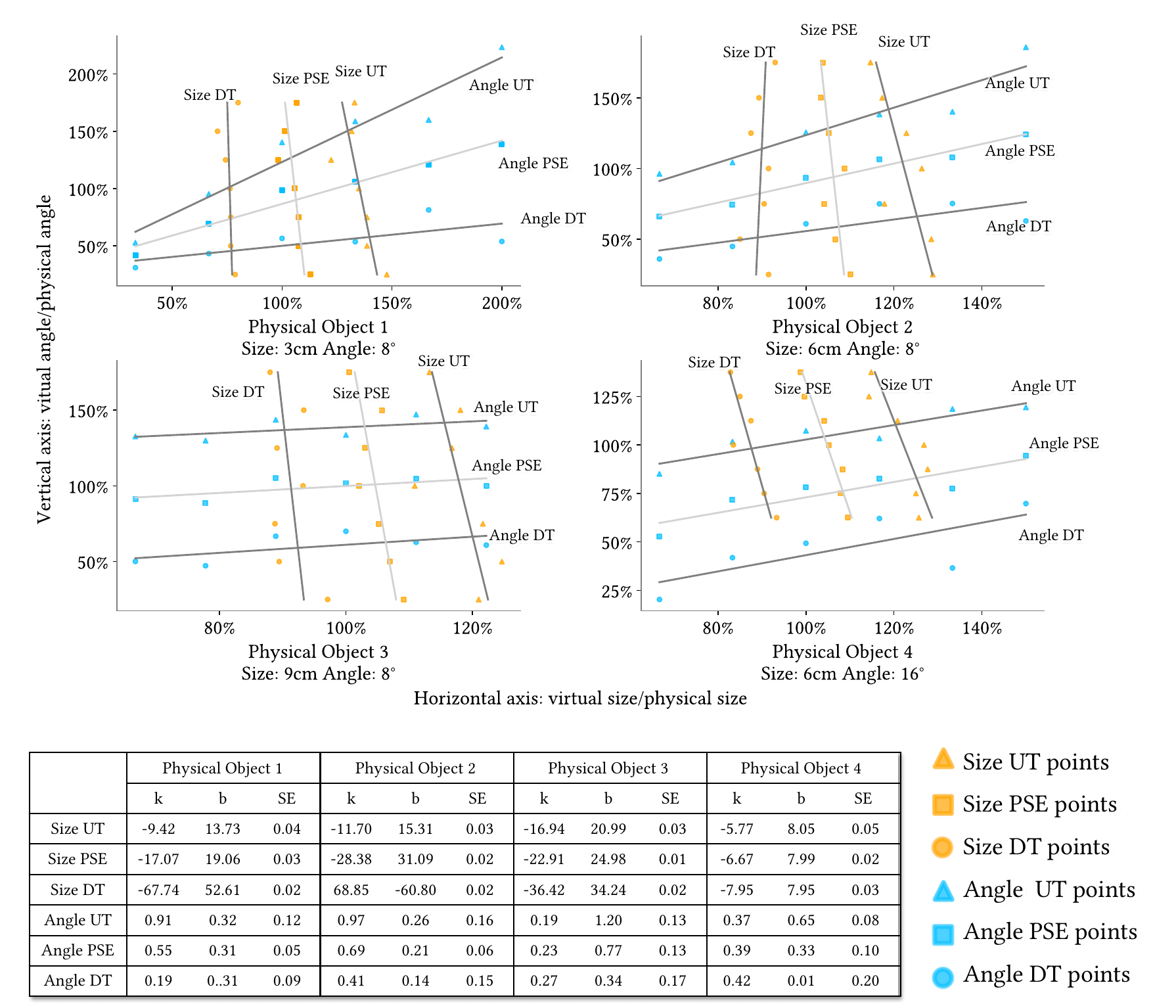}
  \caption{Size perception thresholds and PSEs change with virtual angles (vertical lines in each subplot from left to right: size downscaling thresholds, size PSEs, size upscaling thresholds), and angle perception thresholds and PSEs change with virtual sizes (horizontal lines in each subplot from top to bottom: angle upscaling thresholds, angle PSEs, angle downscaling thresholds). The fit coefficients \(k\) and \(b\), and the standard errors (SEs) are listed in the table for each line.}
  \Description{}
  \label{fig:th2D}
\end{figure*}
For the 3cm-8\textdegree{}, 6cm-8\textdegree{}, 9cm-8\textdegree{} and 6cm-16\textdegree{} physical objects, the thresholds and PSE of size and angle perception are plotted in Fig. \ref{fig:th2D} and we fitted them with linear functions:
\small
\begin{equation}
f(x) = kx+b
\end{equation}
\normalsize
where k is the slope and b is the intercept of the fitted line.

The horizontal axis is the size incongruence (virtual size/physical size) and the vertical axis is the angle incongruence (virtual angle/physical angle). Within the quadrilateral formed by the upscaling thresholds and downscaling thresholds of the size and angle perceptions, the illusion of VR works and it's considered difficult for the users to notice the differences between the physical proxies and the virtual objects. Therefore, we define the quadrilateral as the two-dimensional illusion space of perception in VR. 


\subsection{Three-dimensional Illusion Space: how the perception changes with physical properties}

\begin{figure*}[t]
     \centering
     \begin{subfigure}[b]{0.49\textwidth}
         \centering
         \includegraphics[width=\textwidth]{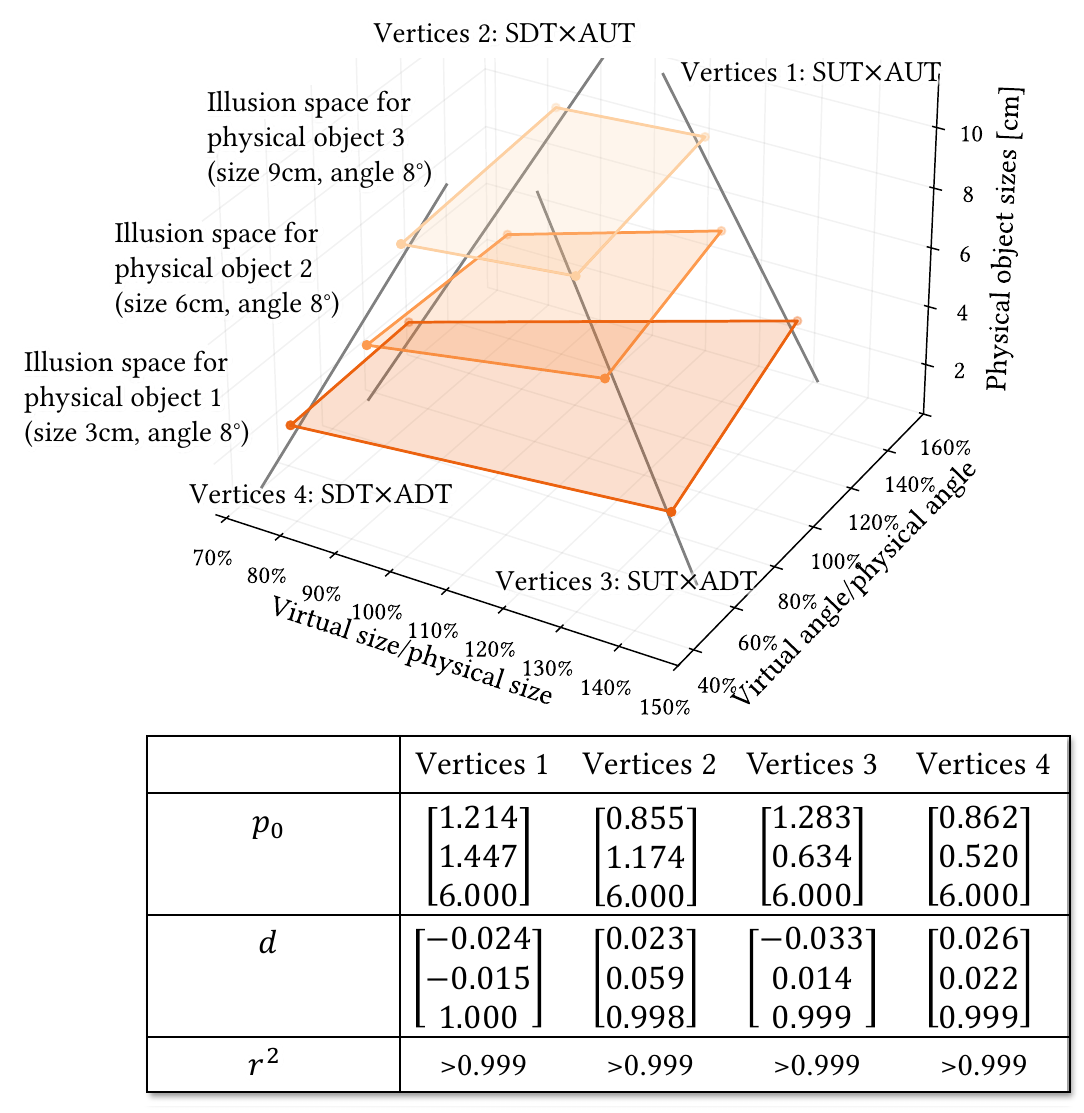}
         \caption{Illusion spaces change with physical sizes. Results are from physical objects 1, 2 and 3 (sizes: 3cm, 6cm and 9cm, angles: 8\textdegree{}}
         \label{fig:sizeillusionspace}
     \end{subfigure}
     \hfill
     \begin{subfigure}[b]{0.49\textwidth}
         \centering
         \includegraphics[width=\textwidth]{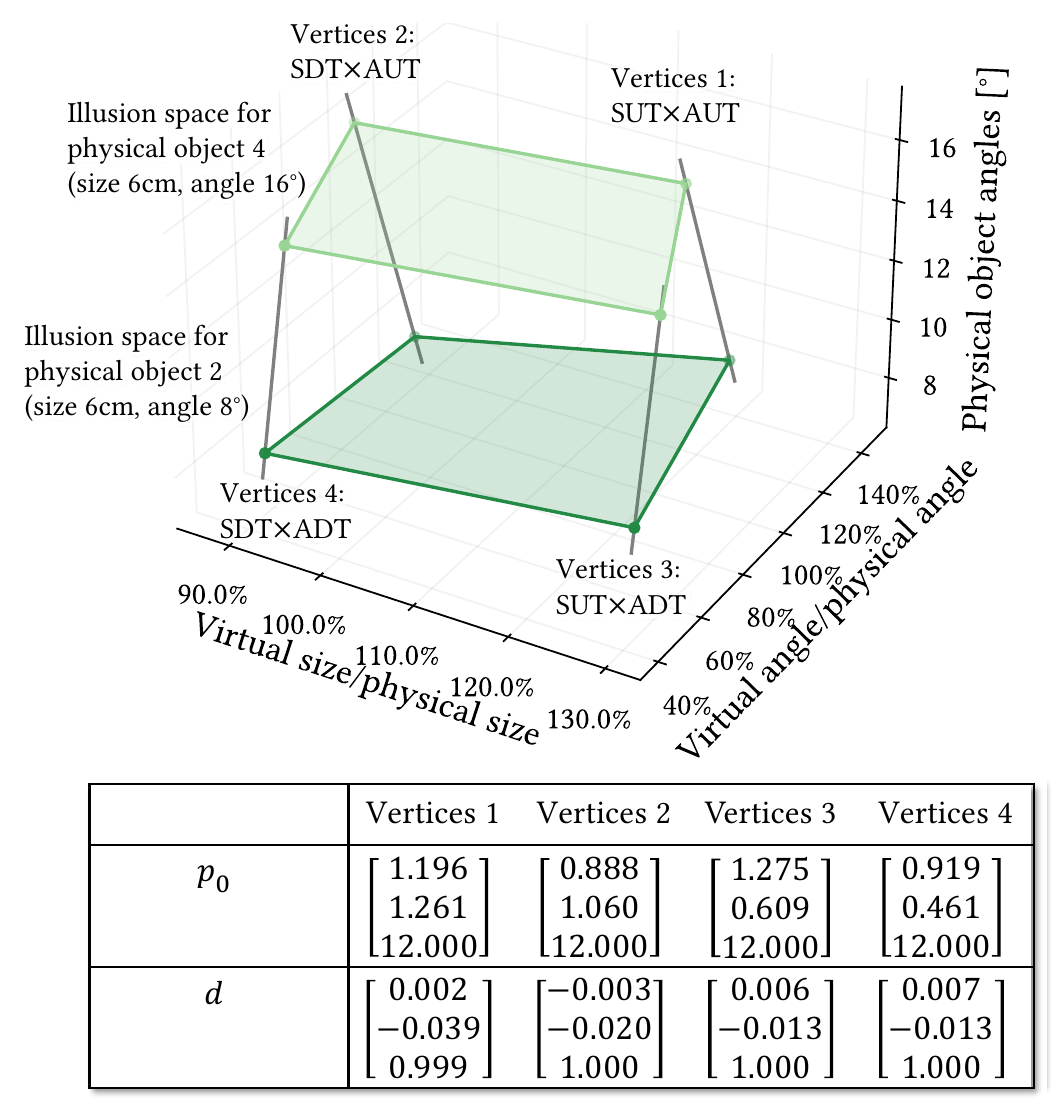}
         \caption{Illusion spaces change with physical angles. Results are from physical objects 2 and 4 (sizes: 6cm, angles: 8\textdegree{} and 16\textdegree{})}
         \label{fig:angleillusionspace}
     \end{subfigure}
     \caption{Results of illusion space changes with sizes (a, left) and angles (b, right).}
     \label{fig:lab}
\end{figure*}



When we add a third dimension (i.e. physical size or physical angle) as shown in Fig. \ref{fig:sizeillusionspace} and Fig. \ref{fig:angleillusionspace}, the figures show the trends of how illusion spaces a physical object can render change with its physical properties. According to our results, we assume the size upscaling threshold (SUT), the size downscaling threshold (SDT), the angle upscaling threshold (AUT) and the angle downscaling threshold (ADT) can be fitted in straight lines and the quadrilateral formed by the four threshold lines is the illusion space for each physical object.
Therefore, the illusion space can also be expressed via four vertices (crossing points of the size and angle perception thresholds), namely \(V_{SUT\times AUT}, V_{SDT\times AUT}, V_{SUT\times ADT}, V_{SDT\times ADT}\). By fitting the same vertex for different physical sizes/angles in a 3D line, we can obtain the mathematical description of how the vertices change with physical sizes/angles, as plotted in Fig. \ref{fig:sizeillusionspace} and Fig. \ref{fig:angleillusionspace}.

The 3D line can be expressed as:
\small
\begin{equation}
    L(t) = P_0+d\cdot t
\end{equation}
\normalsize

where \(L(t)\) is a point on the line for a given parameter \(t\), \(P_0\) is a point on the line, typically the centroid of the points for the fit, \(d\) is the direction vector for the line and \(t\) is a scalar parameter that varies along the line. Therefore, for physical objects 3cm-8\textdegree{}, 6cm-8\textdegree{} and 9cm-8\textdegree{} with different physical sizes, we fitted four lines for the vertices \(L^{size}_{SUT\times AUT}\), \(L^{size}_{SDT\times AUT}\), \(L^{size}_{SUT\times ADT}\) and \(L^{size}_{SDT\times ADT}\).




For physical object 6cm-8\textdegree{} and 6cm-16\textdegree{} with different physical angles, we fitted four lines (link the 2 points on each line) for the vertices \(L^{angle}_{SUT\times AUT}\), \(L^{angle}_{SDT\times AUT}\), \(L^{angle}_{SUT\times ADT}\) and \(L^{angle}_{SDT\times ADT}\). 

For physical objects (1, 2 and 3) with different sizes, the illusion space for size perception becomes smaller when the physical size is larger, while the angle perception isn't obviously influenced. The change of size perception with physical sizes has also shown the same pattern in the resized grasping study\cite{bergstrom2019resized}. On the other hand, the illusion space area of angle perception decreases when the physical taper angle is larger, as shown in Fig.  \ref{fig:angleillusionspace}. \add{Although physical sizes and taper angles are different properties with different units, their influence can be compared in terms of application (i.e. the influence of changing 1 cm on the sizes and changing 1\textdegree{} on the taper angles. According to the results of Fig. \ref{fig:sizeillusionspace} and Fig. \ref{fig:angleillusionspace}, \textbf{changing physical sizes has more influence on the size and angle perception thresholds than changing physical taper angles.}}

Our results show that as the physical size increases, the size of the resultant illusion space decreases. On the one hand, this could be seen as a violation of Weber's law, which states that 
as the intensity of a stimulus increases, so too does the JND (i.e., you need a larger step change in stimulus to perceive a difference)
~\cite{ekman1959weber}. However, Smeets et al. have proposed that the size should not be seen as the property of stimuli to be examined in Weber's law in grasping tasks~\cite{smeets2008grasping}. Instead, according to Smeets et al., for different object sizes the stimuli actually are the finger positions when grasping with simultaneous visual feedback, while the size perception can only be estimated in delayed grasping where the subjects remember the size information. Similarly to \citet{utz2015biomechanical}, we would argue that the stimuli is likely a complex function of joint positions, grip forces, friction, and visual feedback. Our working hypothesis is that the illusion space is smaller for larger items as you are towards the maximum aperture of the finger joints nearer the hand, which provides a clearer perceptual starting point for assessing the size of an object (i.e., 
we know that our own maximum aperture is approximately 12cm, for example, providing a starting point for size perception). 
Understanding and evaluating this hypothesis remains an interesting avenue for future work.

\subsection{Mathematical expression of the illusion space}

\begin{figure*}
  \includegraphics[width=0.95\textwidth]{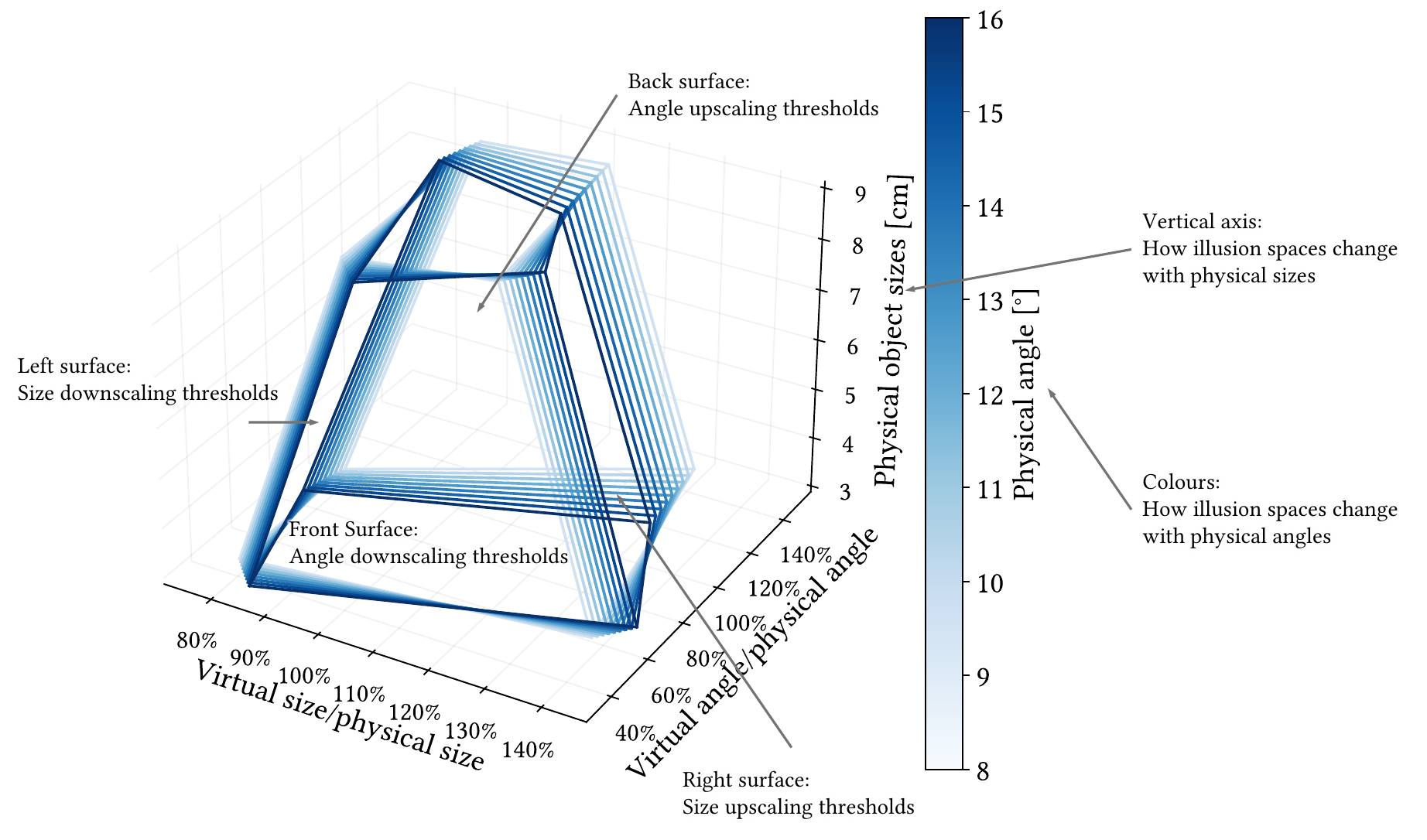}
  \caption{Illusion spaces (including different physical sizes and angles)}
  \Description{}
  \label{fig:illusionspace}
\end{figure*}




\begin{figure*}
  \includegraphics[width=0.95\textwidth]{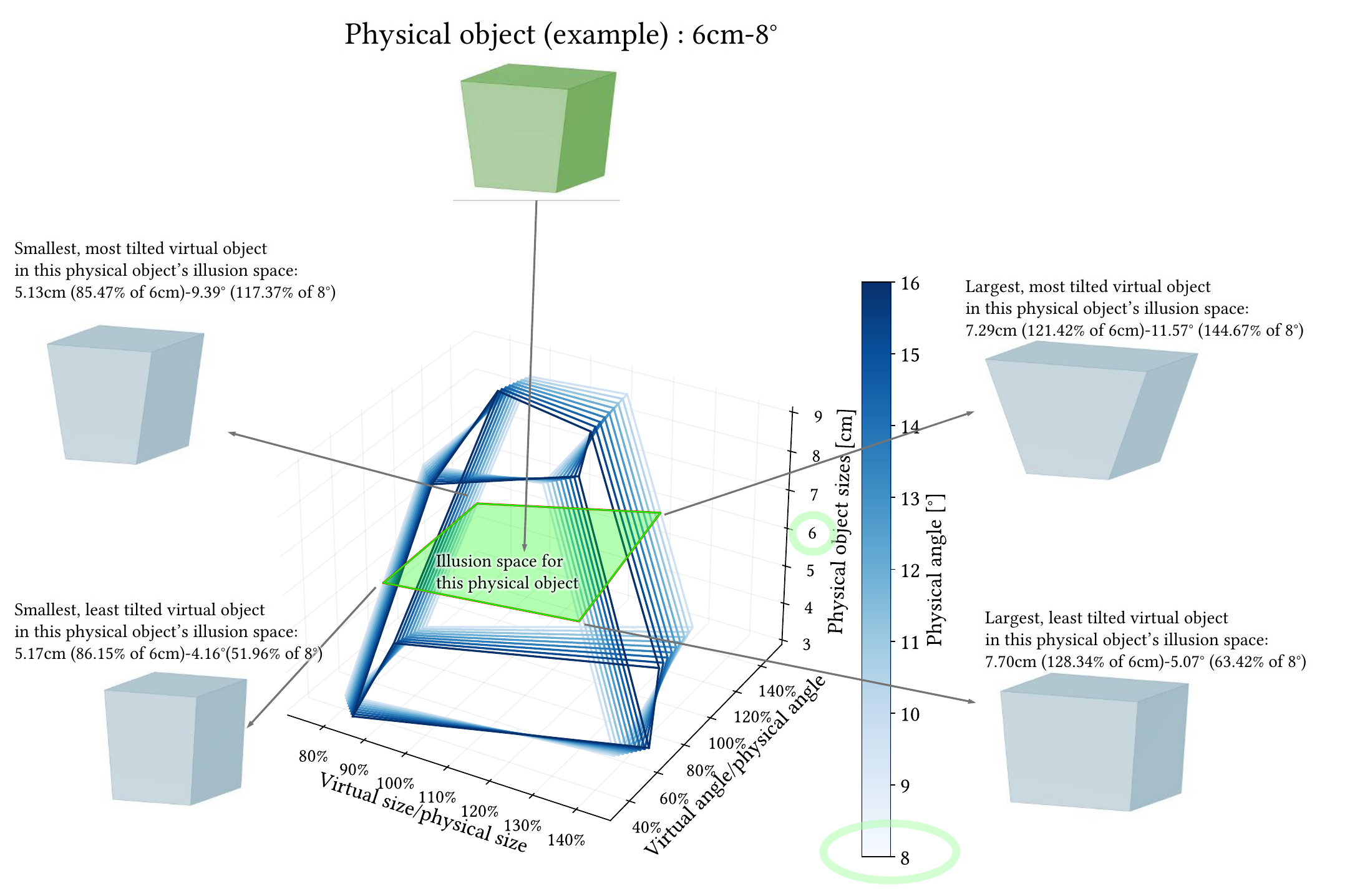}
  \caption{Illusion space for an example physical object (6cm-8\textdegree{}. The vertices are the smallest-least tilted, smallest-most tilted, largest-least tilted and largest-most tilted virtual objects it can represent in VR.}
  \Description{}
  \label{fig:example}
\end{figure*}

If the physical objects within our range of sizes and angles all fit in this pattern, we are able to predict the size and angle of virtual objects that a certain physical object can represent. The mathematical expression can be obtained by combining the four dimensions we discussed before: virtual size, virtual angle, physical size and physical angle. 

Specifically, we calculated how much the four vertices change when the physical angle changes from 8\textdegree{}, and applied this change to the illusion space with virtual size, virtual angle and physical size to complete the four dimensions in the prediction.

For a physical object of size \(S_p\), angle \(A_p\) and virtual angle incongruence \(A_v\)(ratio of virtual angle and physical angle), the downscaling threshold for size perception \(DT_{Size}\) is:
\small
\begin{equation}
    DT_{Size} = \frac{-10A_pA_v+5A_p-2A_vS_p+87A_v+S_p^2+35S_p+275}{-7A_p+37S_p+489}
\end{equation}
\normalsize

Note that for the physical sizes [cm] and the physical angles [\textdegree{}], only the values should be calculated (without their units). The upscaling threshold for size perception \(UT_{Size}\) is:
\small
\begin{equation}
    UT_{Size} = \frac{-4A_pA_v+A_pS_p-30A_p+9A_vS_p-93A_v+S_p^2-86S_p+1778}{-26A_p-29S_p+1197}
\end{equation}
\normalsize

For a physical object of size \(S_p\), angle \(A_p\) and virtual size incongruence \(S_v\)(ratio of virtual size and physical size), the downscaling threshold for angle perception \(DT_{Angle}\) is:
\add{
\small
\begin{equation}
    DT_{Angle} = \frac{A_pS_p-11A_p-S_p^2-8S_pS_v-10S_p+165S_v+275}{-A_p-59S_p+785}
\end{equation}
\normalsize
}
The upscaling threshold for angle perception \(UT_{Angle}\) is:
\add{
\small
\begin{equation}
    UT_{Angle} = \frac{A_pS_p-20A_pS_v+10A_p-S_p^2-74S_pS_v+23S_p+873S_v-54}{5A_p-47S_p+604}
\end{equation}
\normalsize
}
Not only the size and angle perception thresholds are predicted with our illusion space model, but the PSEs can also be calculated by:
\small
\begin{equation}
    PSE = \frac{UT+DT}{2}
\end{equation}
\normalsize

Therefore, we managed to describe the interaction between physical size, physical angle, virtual size and virtual angle in grasping in our illusion space model. According to the mathematical expression, the illustration of the illusion space including these four dimensions is shown in Fig. \ref{fig:illusionspace}, which demonstrates the interaction relationship among the factors. The model was constructed based on the data collected in our user study and we believe it can be applied to larger range of sizes and angles. 

\add{As shown in Fig. \ref{fig:example}, we interpret the illusion spaces with an example physical object of 6 cm and 8\textdegree{}. The green area shows the illusion space of this specific physical object and the vertices are the smallest-least tilted, smallest-most tilted, largest-least tilted and largest-most tilted virtual objects in its illusion space. Other borders of the illusion space can also be calculated from the mathematic expression.}

\add{In order to support designers' use of these equations and findings, we provide a code tool that allows for the easy calculation of the perception thresholds given the properties of physical objects, and as such what virtual objects a physical object can represent. \addfinal{Similarly, the tool enables physical object requirements to be determined from known virtual properties\footnote[4]{https://github.com/JianUnimelb/IllusionSpaceTool}.} }
\section{Discussion}

\add{In this study we estimated how perceptual thresholds are coupled for size and taper angle and built complex, multi-dimensional illusion spaces. In this section, we discuss what we learn from these illusion spaces and how they may be useful to the VR haptics community.}

\subsection{Key Findings from the Illusion Spaces}
For size perception, we generally perceive physical objects to be larger than they are, and virtual size differences of -0.5cm to +1cm from the physical sizes will go unnoticed in most cases. This provides a general starting estimate of a physical object's perceptual limits. From our results of 3cm-8\textdegree{}, 6cm-8\textdegree{} and 9cm-8\textdegree{} in Fig. \ref{fig:size1}, \textbf{PSEs are closer to the true size of the object when the physical objects are large or small in size, than when they are of medium size}. Our work confirms \citet{bergstrom2019resized}'s findings that the ranges of the illusion spaces decrease with the increase of physical sizes. This suggests that, towards the maximal extent of our grasping range, we are more sensitive to objects' size, than when our fingers are closer together and grasping smaller objects.

Taper angles are generally underestimated and objects will be perceived as less titled than they actually are. When sizes are congruent (between physical and virtual objects), an angle incongruence of $\pm2$\textdegree{} from the physical angle is typically not noticeable. Again, these could form starting estimates for an object's perceptual limits. \textbf{PSEs are closer to the physical object's taper angles when objects are smaller or larger, than when they are medium-sized.}

Additionally, \textbf{as the physical taper angle increases, we need greater changes in order to perceive any difference (the JND increases),} while the PSE for size perception gets closer to the true size of the object, as shown in the results of the 6cm-0\textdegree{}, 6cm-8\textdegree{} and 6cm-16\textdegree{} objects in Fig. \ref{fig:size1} and Fig. \ref{fig:angle1}.

For the angle perception in Figure \ref{fig:th2D}, the thresholds and PSE all show an increasing trend when the virtual sizes are larger, which means \textbf{as the virtual sizes increase, we perceive that the taper angle increases and it becomes more difficult to notice the angle incongruence.} On the other hand, for size perception, the thresholds and PSE also show trends with the change of virtual angles. \textbf{When the virtual angle decreases, we increasingly overestimate the size of the physical object. 
} These facts show that an interplay between properties and incongruences of physical and virtual objects exists. 
\add{By examining the slope of the lines, in all these cases \textbf{\addfinal{altering virtual sizes changes the angle perception thresholds more than angle changes impact size perception}.}}
These results suggest a new method to expand the ability for physical objects to represent virtual ones, by changing other virtual properties instead of changing the physical objects.


\subsection{Implications for VR Haptics}
This multi-dimensional illusion space model offers valuable knowledge for haptics designers in VR. When designing haptic devices with similar properties to our settings, given the physical size and angle, we can immediately obtain the range of virtual objects with various sizes and angles this physical object can represent. 

\add{For haptic controllers and physical passive proxies, smaller designs have the advantage of rendering a virtual object with a larger range of sizes (proportional to the controller sizes), according to our findings. For angle perception, devices with larger physical angles are likely to be perceived as showing less tilted angles in VR.}

Our model also offers brand-new options in haptic feedback designs by showing how virtual factors can interplay with each other and influence perception. For instance, with increases of virtual sizes, the perceived angle of the haptic feedback increases. On the other hand, with the increase of virtual angles, the perceived size becomes smaller. The model demonstrates that a user's haptic perception can be altered by changing another interplaying virtual factor. In turn, this can serve to expand the feedback range of an active controller, without any changes to its hardware design.

\add{For more complex or different haptic devices, we have provided insights into how size and angle perception may influence each other. For example, in the controller design of \citet{ulan2024}, the shape of virtual objects can be dynamically and precisely displayed by the controller within its mechanical limits, but our illusion space models show how the range of feedback can be largely increased through illusions. Similar benefits can be found for hand-mounted devices, too, such as Wolverine~\cite{choi2016wolverine} and Claw~\cite{choi2018claw}.}

\add{Through these interplay effects, we have revealed a greater haptic coverage for passive proxy objects (in terms of illusions of similarity~\cite{propping}). Where prior work has explored \textit{spatial} haptic coverage\cite{clarence2024stacked}, we begin to contribute deeper insights into the broader opportunities of \textit{geometric} haptic coverage.}

\add{Our tool for identifying illusion spaces for passive objects further improves designers' and users' ability to represent more virtual objects with the same, or even less, physical props. Building on the concepts of Substitutional Reality~\cite{simeone2015substitutional}, for example, our results increase the potential use of objects around the home within VR and may serve to further reduce design constraints when producing physical props~\cite{zhu2019haptwist}.}



\subsection{Limitations and Future Work}
To control variants in the experiment, the same 2AFC questions were presented to the participants in every task. Although we've analysed data from most tasks, the resultant proportions of 6cm-0\textdegree{} physical object cannot be fitted to a sigmoid function and show a different pattern from our other results. This different pattern stems from the fact that the haptic feedback was always less tilted (other than when it exactly matched) and this disrupted the 50\% probabilities in 2AFC results (the ambiguous point where haptics and visual feedback are considered to be perceived as the same). However, we believe it's still important to study the perception of virtual angles around both sides of a straight object (0\textdegree{} taper angle). Studying these objects will require a different formulation of the question or, perhaps, a different psychophysics method. Alternatively, other biometric methods are emerging that leverage gaze tracking and EEG to determine perceptual thresholds~\cite{feick2024predicting}, and they could prove valuable here. This remains an interesting avenue for future work.


While we highlight that prior studies have only concentrated on a single factor, our study is constrained to size and taper angle. \add{There remain many other factors, some of which have been studied solely but we expect likely interplay with other factors in interesting ways should be explored and added into the illusion space. Surface texture, for example, has influence on the friction cones in grasping according to the grasping model~\cite{zheng2022human} and thus may influence the angle perception. However, its influence on pressure magnitude and force direction may have a lesser influence on the size perception limits. Mass, on the other hand, has been considered related with size perception in many previous studies~\cite{cross1975relation,schmidtler2017influence} and can be included as another factor in the illusion spaces. Mass likely impacts joint torques when lifting, which goes beyond the angular joint configurations that we have been largely concerned with here. Curvature, conversely, is a further dimension of shape that could perform similarly to taper angle as its effects on finger span and friction should be limited.}

\add{Apart from factors of the grasped objects, the grasp types can also alter the limits of the illusion spaces. We chose to examine the Palmar Pinch as a popular choice in human computer interaction (HCI) (e.g., ~\cite{kim2022spinocchio,ulan2024}, but other grasp types could easily result in quite different joint orientations, torques, and force profiles. From the GRASP taxonomy\cite{feix2015grasp}, for example, grasps such as Large Diameter, Power Sphere and Sphere Finger include the palm in grasping and the force distribution changes dramatically, with the fingers largely acting as stabilisers into the palm, highly likely resulting in changing of perception limits. When the thumb is adducted in grasping (e.g. Adducted Thumb and Fixed Hook), or when more fingers are included in grasping (e.g. Quadpod and Precision Sphere) the fingers collaborate in a different way which will change the perception limits estimated in our illusion spaces. Each of these properties and grasp type has a unique impact on the biomechanical and sensory factors of grasping and further studies would serve to deepen our knowledge of the underlying mechanisms at work here. With further studies, we may begin to look at generalisable, biomechanical models of grasping that reveal much deeper insights into perceptual thresholds and object selection.} 
\section{Conclusion}
Previous studies have estimated perception thresholds of a single factor in grasping. In this paper, we proposed a model of illusion space of size and angle perception in VR. Instead of single-factor threshold estimation, we explored the interplay among four factors: physical sizes, physical angles, virtual sizes and virtual angles with the model. We conducted 2AFC experiments with multiple combinations and summarised the pattern of the results with mathematical expressions and figure illustrations. We reveal a multi-dimensional \textit{illusion space} for both size and taper angle. 

The physical sizes and taper angles have obvious influences on the perception. When physical objects are small, they can represent a greater variety of virtual object sizes in VR. A more tilted physical taper angle allows these objects to mimic a broader spectrum of virtual sizes and angles. 
In addition to physical characteristics, we found that increasing virtual object size substantially enlarges the perceived angle and expands the range of angles a physical object can simulate. On the other hand, increasing the virtual angle leads to a smaller perception of size, illustrating a strong interaction between size and angle in VR-induced illusions.

We introduce a function that better captures a physical object's haptic potential, highlighting how size and taper angle work together to create illusions. Moreover, we explore how these insights could be applied to the design of haptic systems, advancing the development of high-resolution haptic devices.
 We provide a model that allows for the prediction of perception thresholds. Our findings broaden our understanding of how to design for detailed and accurate haptic feedback in VR.

\begin{acks}
  We gratefully acknowledge the support provided by the Australian Research Council Discovery Early Career Research Award (Grant No. DE210100858) and the support of the Melbourne Research Scholarship from University of Melbourne.
\end{acks}
\bibliographystyle{ACM-Reference-Format}
\bibliography{main}









\end{document}